\documentclass[aps,prl,twocolumn,groupedaddress]{revtex4}
\usepackage{graphicx}
\begin{document}
\title{Alignment of chiral order parameter domains in Sr$_{2}$RuO$_{4}$ by magnetic field
cooling}
\author{Fran\c{c}oise Kidwingira}
\email[ ]{kidwingi@stanford.edu}
\altaffiliation{Current address: Geballe Laboratory for Avdanced Materials, Stanford University.}
\author{Joel D. Strand}
\author{D. J. {Van Harlingen}}
\affiliation{University of Illinois at Urbana-Champaign}
\author{Yoshiteru Maeno}
\affiliation{Department of physics, Kyoto University, Kyoto 606-8502, Japan}

\date{\today}

\begin{abstract}

Superconductivity in Sr$_{2}$RuO$_{4}$ is unconventional, believed to
be of $p_{x}\pm ip_{y}$ pairing symmetry. These two degenerate
order parameters allow the formation of chiral domains separated by
domain walls. In a Josephson junction formed on the edge of a
single crystal of Sr$_{2}$RuO$_{4}$, the chiral domains can create a variation of the phase in the tunneling direction causing
interference effects which suppress and modulate the critical current of the
junction. Cooling the junction in a magnetic field lifts the
degeneracy between the order parameter states and induces a preferential chirality, significantly modifying the
phase interference. We present experimental results on
Sr$_{2}$RuO$_{4}$/Cu/Pb Josephson junctions cooled in a magnetic field
showing a dramatic enhancement of their critical current, giving direct evidence for the presence of chiral domains and their alignment in a magnetic field.
\end{abstract}

\maketitle

Sr$_{2}$RuO$_{4}$ (SRO) was discovered to be superconducting in 1994\cite{SROdisc}. Since
then it has proven to be a very intriguing and complicated superconductor. A large
body of evidence \cite{MaenoMackenzie} points towards unconventional
superconductivity with an order parameter of the form $p_{x}+ i
p_{y}$: the electrons are paired in an odd orbital, spin triplet
channel that breaks Time Reversal Symmetry (TRS). The broken TRS,
first observed by muon spin rotation\cite{Luke1998} and later
verified by polar Kerr effect \cite{Xia2006}, is especially
interesting since it can lead to the appearance of order parameter
domains of opposite chirality $p_{x}\pm i p_{y}$ in the material, much like in a ferromagnet.
 Texture of the order parameter, has been observed in the Helium 3 A-phase, which is the superfluid
analog of SRO, but until recently domains were never directly observed
in a superconductor. They have been proposed to explain unusual features in the ultrasound attenuation and
the rate of vortex creep dynamics in superconductors thought to break TRS: UPt$_{3}$, UBe$_{13}$ and SRO
(\cite{SigristUeda},\cite{MotaUPt3},\cite{MotaUThBe13},\cite{MotaSRO}),
but their first unequivocal observation was through Josephson
interferometry experiments done on SRO \cite{kidwingira}. The
presence of order parameter domains opens a whole new field in superconductivity and
efforts are being made to understand domains nucleation and dynamics. Many of their
properties are still not understood. For example, scanning SQUID and
Hall microscopy have repeatedly failed to image the domains
(\cite{moler1},\cite{moler2}) although theory predicts magnetic fields that are well above the resolution of these instruments.

In this Letter, we aim to further that understanding by manipulating
the domains using applied magnetic fields. We find that field
cooling increases the probability of one type of domain
which translates into a critical current much higher than that
observed in the same junction cooled in zero field. We also found an
unexpected memory effect with training characteristics.

 Superconductors with broken TRS were first studied theoretically by Volovik and Gor'kov \cite{Volovikletter}
who found that they can have unique magnetic properties associated
with the variation of the phase of the superconducting order
parameter. Since the phase of a $p_{x}\pm ip_{y}$ order parameter
continuously varies from $0$ to $2\pi$ in $k$-space, a spontaneous current will flow within a coherence length of the surface of the material even in the ground state. To cancel the magnetic field in the bulk of the superconductor, additional Meissner screening currents flow within a
penetration depth of the surface. Thus, although there is no net magnetization, there should still be a magnetic moment at the surface as
well as at sites with a suppressed order parameter such as defects
and impurities. To offset the cost of magnetic energy, the discrete degeneracy of order parameters may promote the formation of order parameter domains similar to those in ferromagnetic materials \cite{Volovikletter}, but they may also nucleate spontaneously and become trapped by pinning of domain walls at defects or impurity sites. These
so-called chiral domains are defined by the two directions of phase
winding associated with the degenerate order parameters
$p_{x}+ip_{y}$ and $p_{x}-ip_{y}$ and they couple to applied magnetic
fields (\cite{Volovikletter}, \cite{Volovik}, \cite{Sigrist1989}) via the spontaneous currents. The magnetizations associated with the two chiral domains have
opposite signs so that an externally applied field causes a difference in
their free energies, thus lifting the degeneracy of the order
parameters. The chiral domains with a magnetization parallel to the
applied field increase in size while the others shrink.

\begin{figure}
    \centering
    \includegraphics[width=8cm]{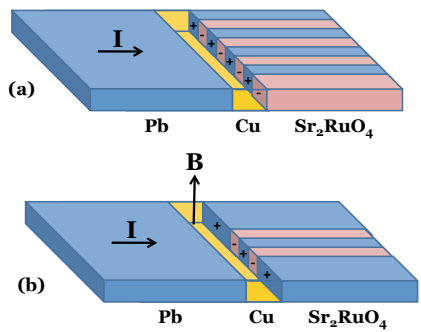}
    \caption[Cartoon of junction]{ Cartoon picture of the junction geometry for a phase difference of $\pi$ between neighboring domains. (a) The supercurrent flows in the indicated direction I in the blue domains and in the opposite direction in the pink domains. (b) The applied magnetic field B favors one of the two order parameters and increases the net current through the junction.
 \label{fig:domaincartoon}}
\end{figure}

Josephson junctions have largely contributed to the understanding of the material. The determination of allowed and forbidden tunneling directions established constraints on the order parameter\cite{Jin}, SQUID interferometry showed the orbital part of the order parameter to be of odd symmetry\cite{Nelson}, and they have proven to be very sensitive probes to the presence of domains (\cite{kidwingira},\cite{Kambara}). Here, we use their critical current and
interference patterns as a metric for the number and size of
order parameter domains as we cool the junctions in a magnetic field
parallel to the c-axis of the crystal. For a junction on the side of
a crystal incorporating many domains, the phase difference across the junction can change from one domain to the next causing interference effects that reduce its critical current. The most drastic case is that of a phase difference of $\pi$ where the currents in neighboring
domains flow in opposite directions and the measured critical
current is the net sum of all of them (Fig.\ref{fig:domaincartoon}). Cooling the crystal in a
magnetic field changes the ratio of domains, reducing the amount of cancellation.

\begin{figure}
    \includegraphics[width=8cm]{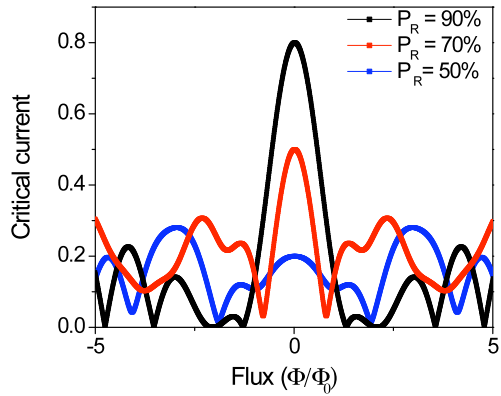}
    \caption[Simulation of diffraction pattern for different probabilities for the domains]{Simulation
    of diffraction patterns of a junction where the percentage of chiral domains that are right-handed
    is $50\%$, $70\%$ and
    $90\%$ as shown on the figure. As the probability of one type of domain increases, so does the critical current across the junction.

    \label{fig:simFC}}
\end{figure}

As one
domain type increases in area, the critical current of the junction
measured at zero magnetic field correspondingly increases and its
diffraction pattern is modified, tending towards a Fraunhofer interference pattern. To illustrate
this behavior, we performed computer simulations of critical current
modulations in an applied magnetic field when one type of domain
increasingly becomes more probable than the other. Here, $P_{L}$ and
$P_{R}$ are the percentage of left- and right-handed chiral domains
in the material. The calculation is done for tunneling from a
superconductor with an s-wave order parameter to one with a
$p_{x}\pm ip_{y}$ order parameter for a junction with $10$ domains
of random size across its width and assuming a phase difference of $\pi$ between the two polarities of chiral domains. The critical current is normalized:
$I_{c}=1$ is the maximum current that would flow through a domain
free junction at zero applied field. The results are given in
Fig.\ref{fig:simFC}.
As $P_{R}$ increases, the magnitude of the current increases and a central peak emerges,
reminiscent of the Fraunhofer pattern expected for a domain-free
junction. Of course, the detailed shape of the pattern depends on the size and number of domains that are used in the
calculation but the qualitative trend is clearly shown by the simulations. 

We performed the experiment, using the magnitude of the critical
current  and its magnetic field dependence as a measure of domain alignment. The high quality single
crystals used in the experiment were grown using the floating zone
method as described elsewhere \cite{MaoCGrowth}. The crystals were
glued onto a glass substrate and the junctions defined using a flexible membrane mask. They are parallel to the c-axis, about 200 $\mu$m wide and 50$\mu$m high. 
The samples were then loaded into a vacuum chamber where they were cleaned by
ion milling and then 10nm of Cu and about 1 $\mu$m of Pb were deposited by thermal evaporation. The critical current was measured using a feedback technique where the current output to the sample was automatically adjusted to maintain a set voltage just above the supercurrent regime. We used a SQUID
potentiometer circuit to measure the small voltages generated. The magnetic field was
applied parallel to the c-axis using a Helmholtz coil. The measurements were done in a
$^{3}$He refrigerator at temperatures between 340mK and 1.3K. The data
presented here was taken around 1K.  The magnitude of the critical current vary with temperature but the shape of the observed diffraction patterns is relatively independent of temperature. In between
cooling cycles, the samples were heated to 10K, which is above the transition temperatures of both superconductors in the
device. The samples were then cooled in zero field to 4K to avoid
trapping vortices in the Pb film. Then the field was turned on and
they were cooled below the transition temperature of the SRO
crystal.

\begin{figure}
    \centering
    \includegraphics[width=8cm]{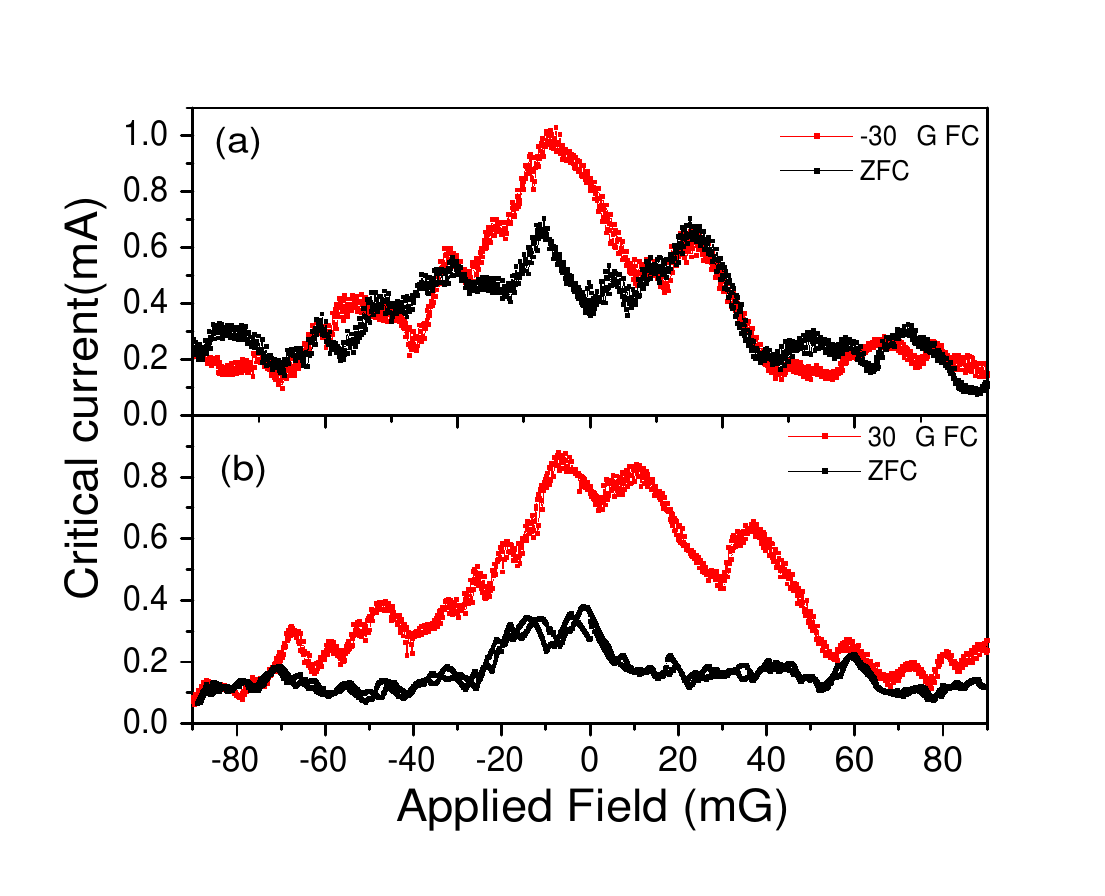}
    \caption[Critical current enhancement by field cooling]{Diffraction patterns for Josephson junctions cooled in (a) positive and (b) negative magnetic field compared to the pattern from the zero field cooling cycle that directly preceded it. The critical current is enhanced for both field polarities. The difference in the two zero field cooled patterns comes from the memory effect discussed in the text.}
    \label{fig:Icenhance}
\end{figure}

This is a challenging experiment because when a junction is cooled
in a magnetic field, vortices can be trapped in the superconductors near the junction that suppress the critical current and 
distort the diffraction patterns. Hence a field cooled junction will
show the combined effects of domain alignment and magnetic
field from trapped vortices; depending on whether and where the vortices are trapped, the
critical current enhancement might not be observable. Thus we performed the experiment at various field values in the mG range but found that the most significant effects were achieved at fields in the 10 to 100$\mu$G range.  We attribute these effects to domain alignment. Significant critical current enhancement was observed for
four samples for both positive and negative fields.
Fig.\ref{fig:Icenhance} shows the diffraction patterns of zero field
cooled (ZFC) and field cooled (FC) junctions for a field of 30$\mu$G at which the maximum enhancement was observed.
The diffraction patterns are not of the Fraunh\"{o}fer type expected
for a domain-free junctions: the domains cause phase interference
that induces distortions \cite{kidwingira}. Even so, a dramatic
enhancement of the critical current by more than a factor of 2 and the emergence of a central peak can be observed, both evidence for less interference in the junction. The change in the shape of the diffraction pattern  is particularly important as it mostly rules out explanations based on a magnetic field induced increase in coupling between the singlet (Pb) and triplet (SRO) superconductors which would change the magnitude of the critical current but not the interference pattern of the junction. The field cooled junctions are never found to be  domain free but this is not surprising given the small fields applied. Out of the four samples measured, the one presented here
showed the most significant effect, and the smallest increase observed
was ~40$\%$. We also cooled some of the samples in a field parallel
to the $ab$ plane. No critical current enhancement was observed, additional evidence that the enhancement in perpendicular field is due to 
order parameter domains in SRO.

 Although the data shows the critical current enhancement,
in agreement with the theoretical predictions, some aspects of observed enhancement are surprising.
First, the optimal enhancement is achieved at extremely small fields, comparable or even smaller than the residual background fields in our system. We are able to observe these effects because the measurement setup has a double layer of mu-metal shielding and a superconducting Pb can that reduces the earth's magnetic field to about 100$\mu$G on average, although this can be significantly lower in the junction area and may not have a large component in the relevant field direction. We should note that the effective field in the junction area is always higher than the nominally applied field because of flux focussing: large areas surrounding the junction are covered by superconductors that expel magnetic field and focus it into the normal regions of the sample. The magnetic field in the junction is self calibrated because the periodicity of the diffraction pattern corresponds to threading one flux quantum through the junction. Although the patterns presented here are complicated, the smallest modulation period corresponds to the largest area, which is the magnetic cross-section of the  entire junction.  Knowing the junction area, we can then calculate the effective magnetic field. For the sample shown above, we estimate the flux to be focused by a factor of 4: the 30$\mu$G externally applied magnetic field value quoted above should translate into about 120$\mu$G in the junction area. Despite the rather small value, the fact that we observe comparable critical current enhancement for both positive and negative applied field indicates that we are indeed applying a net field, and not offsetting a residual background field.
One would think that a larger field would  split the degeneracy more, eventually yielding a domain free junction but we do not observe this and attribute this to the trapping of vortices near the junction which distort the diffraction pattern.
  
\begin{figure}
    \centering
    \includegraphics[width=8cm]{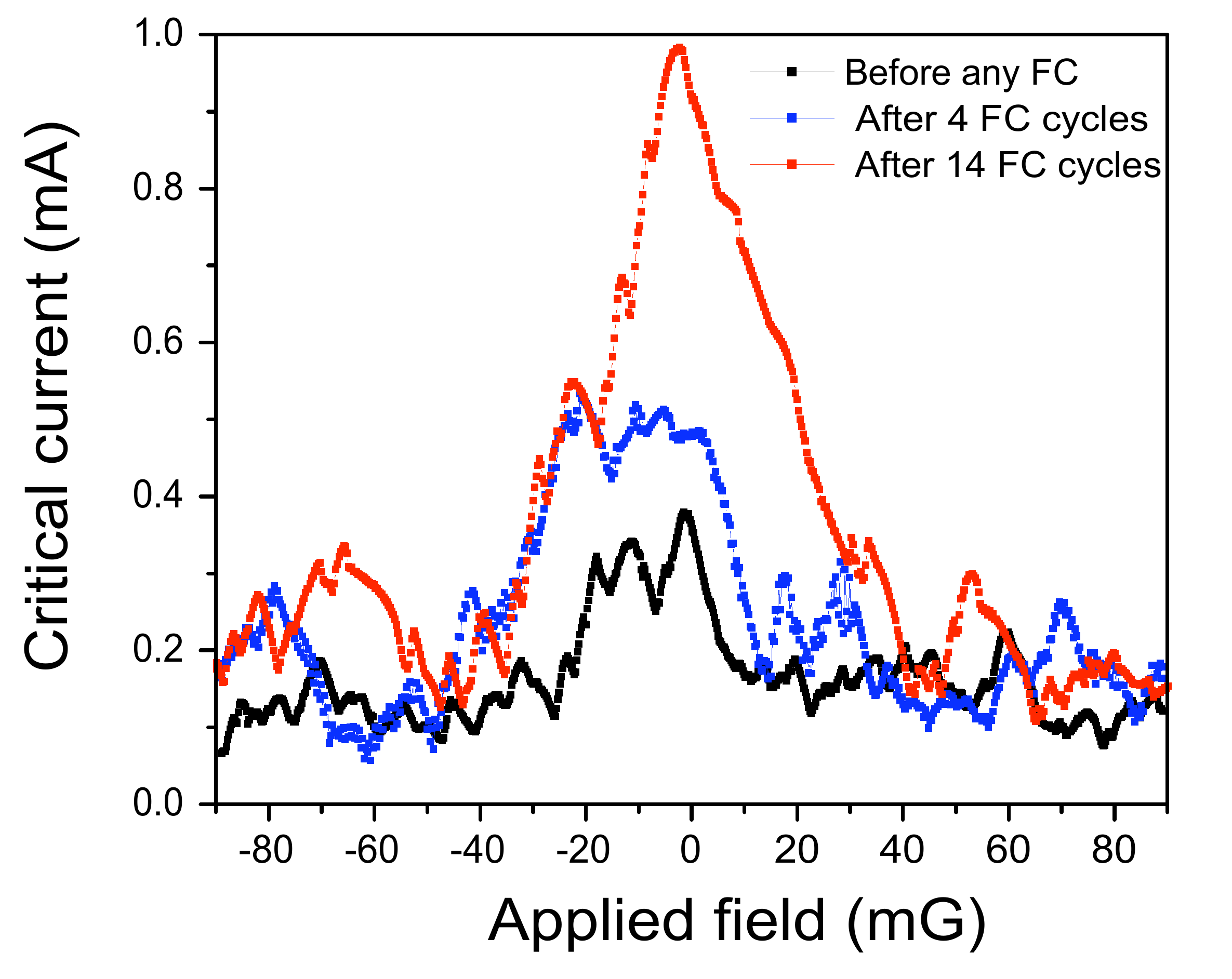}
    \caption[Training of the memory effect]{Gradual critical current enhancement. The critical current increases with the number of field cooling iterations, reminiscent of the training effect in systems with an interface between a soft and a hard magnetic material.
     }
    \label{fig:training}
\end{figure}

\begin{figure}
    \includegraphics[width=8cm]{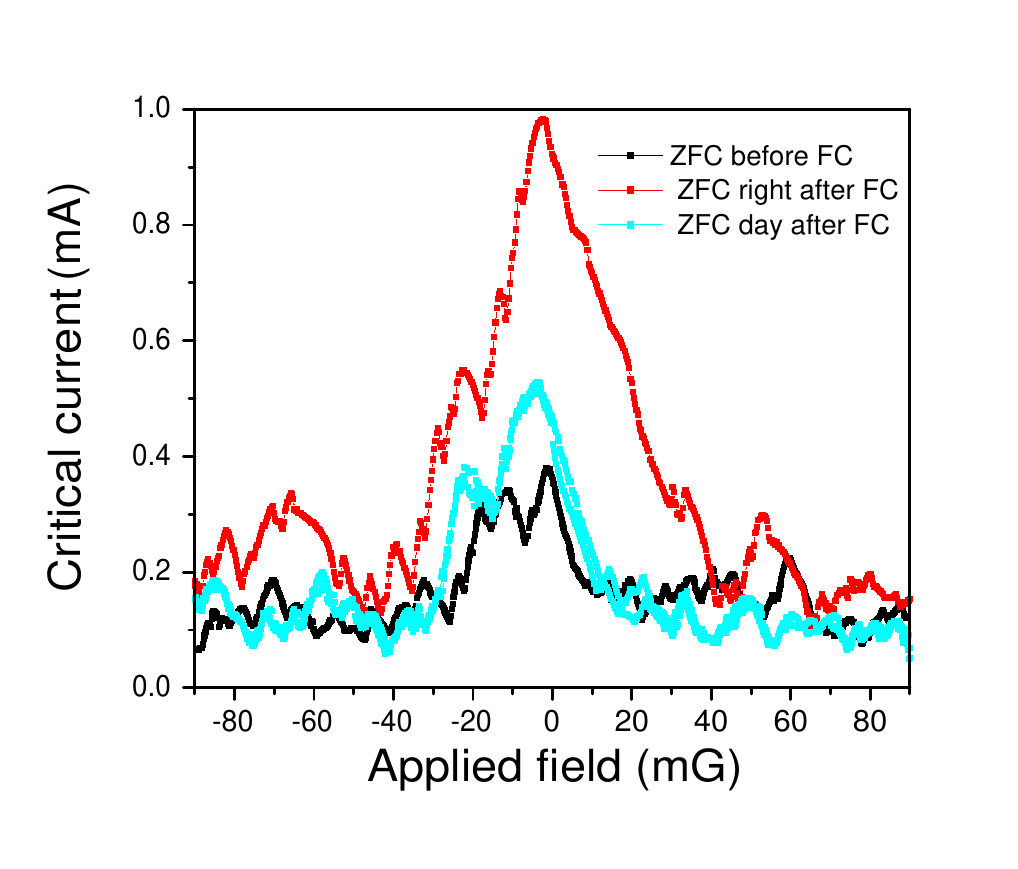}
    \caption[Relaxation of the memory effect]{Relaxation of the memory effect with time and temperature.
    Diffraction patterns of the zero field cooled junction for different cooling cycles: before any field cooling (black),
    right after a field cooling cycle with large $I_{c}$ enhancement (red) and one day after field cooling measurements (blue).}
    \label{fig:memorynextday}
\end{figure}

Second, we observed anomalous training and memory effects in field-cooled samples. The critical current enhancement is found to increase with the number of field cooling cycles, reminiscent
of the training effect observed in ferromagnet/antiferromagnets
multi-layers. The critical current does not reach its peak level in the first iteration of field cooling. Instead, it gradually increases with the number of field coolings, eventually reaching a maximum as can be seen on Fig.\ref{fig:training}. The three curves shown were measured on ZFC junctions, following the respective number of field cooling noted on the figure. We also observed a memory effect: once the critical current of the junction is enhanced,
it retains a high value even when warmed above the transition
temperatures of all superconductors in the device. These observations are plotted in
Fig.\ref{fig:memorynextday}.
The system relaxes with time and
temperature, but at temperatures as high as 77K and times as long
as 24h, the memory of previous cycles is not fully erased. Fig.\ref{fig:memorynextday} shows
3 diffraction patterns taken on a ZFC junction. The first one was
taken before any field cooling (black), the second in a cooling
cycle immediately following the maximum $I_{c}$ enhancement, and the third
was taken one day after the critical current enhancement is
observed. The junction warmed to 77K overnight; the critical
current is lower than immediately after the field cooling but still higher
than in the original state. Because of the long times required to map the junction behavior, we have not determined the conditions
under which the field cooling effects disappears completely.  

The memory effect is obviously not caused by the superconductivity in
the material since it survives well into the normal state. The most likely explanation would be the presence of isolated ferromagnetic regions at the edge of the junction that would become magnetized during the field cooling and then provide a local magnetic field on the subsequent cooling cycles. A ferromagnetic material far away from the junction wouldn't provide a credible explanation because magnetic fields larger than the ones used for field cooling are routinely applied to the sample when measuring diffraction patterns and don't result in a critical current enhancement. Hence whatever causes the memory effect probably originates from inside the superconductor which is shielded by Meissner screening during the diffraction pattern measurements. Magnetic impurities are very unlikely as the superconductivity is very fragile and vanishes with small amounts of impurities, even non-magnetic ones. A transition temperature of 1.5K is only achieved in extremely clean samples. A more likely scenario would be a very small amount of impurity phase of SrRuO$_{3}$, which is a ferromagnet with a Curie temperature of 150K. Another possibility is that of surface magnetism.

In conclusion, we present experimental results showing a dramatic
enhancement of the critical current in SRO/Cu/Pb Josephson junctions
cooled in a magnetic field, which is expected to occur in the
presence of order parameter domains in the SRO crystal. This
increase of the critical current indicates that one of the order
parameters, e.g. $p_{x}+ip_{y}$, is increasing in size while the
other is decreasing, resulting in an increased net current through
the junction. We also report memory and training effects in the field
cooled SRO samples that have yet to be fully explained but may result from local magnetic inclusions.
This work was supported by the
DOE, Division of Materials Sciences under grant DE-FG02-07ER46453 (FK, DVH) and  the NSF under grant DMR-07-05214 (JDS), and through the Frederick
Seitz Material Research Laboratory at the University of Illinois at Urbana-Champaign.

\bibliography{AlignmentDomains}

\end{document}